\documentclass{article}

\usepackage{geometry}
\usepackage{amsmath,amssymb,amsfonts}
\usepackage{amsthm}
\usepackage{microtype}
\usepackage[colorlinks,allcolors=blue]{hyperref}

\usepackage{xcolor}
\definecolor{keywordcolor}{rgb}{0.1, 0.2, 0.6}    
\definecolor{tacticcolor}{rgb}{0.1, 0.2, 0.6}    
\definecolor{commentcolor}{rgb}{0.4, 0.4, 0.4}   
\definecolor{symbolcolor}{rgb}{0.0, 0.0, 0.0}
\definecolor{sortcolor}{rgb}{0.1, 0.2, 0.6}    
\definecolor{attributecolor}{rgb}{0.7, 0.1, 0.1} 

\usepackage{listings}

\usepackage{upgreek}

\title{A Proof-Producing Compiler for Blockchain Applications}

\author{Jeremy Avigad, Lior Goldberg, David Levit, Yoav Seginer, Alon Titelman}

\begin{document}

\maketitle

\begin{abstract}
  \emph{CairoZero} is a programming language for running decentralized applications (dApps)
  at scale. Programs written in the CairoZero language are compiled to machine
  code for the Cairo CPU architecture and cryptographic protocols are used to
  verify the results of execution efficiently on blockchain.
  We explain how we have extended
  the CairoZero compiler with tooling that enables users to prove,
  in the Lean 3 proof assistant, that compiled code satisfies
  high-level functional specifications.
  We demonstrate the success of our approach by verifying primitives
  for computation with the secp256k1 and secp256r1 curves over a large finite field
  as well as the validation of cryptographic signatures using the former.
  We also verify a mechanism for simulating a read-write dictionary data structure in
  a read-only setting.
  Finally, we reflect on our methodology and discuss some of the benefits of our approach.
\end{abstract}





\raggedbottom
\section{Introduction}
\label{sec:introduction}

A \emph{decentralized application} (dApp) is a program that runs on a distributed
ledger system like blockchain. dApps can be used to implement
smart contracts, voting systems, and auctions without requiring oversight by
a centralized institution. Once deployed, a dApp runs
autonomously, its effects registered in a distributed manner by those validating
the blockchain. Running a dApp can be both slow
and expensive, however, because it requires everyone validating the chain to carry out the
computation and reach consensus on the result.

\emph{Cairo}\footnote{https://www.cairo-lang.org/} is a programming language
designed by StarkWare Industries to support running dApps at scale.
Programs written in the Cairo language are compiled to machine
code for the Cairo CPU architecture \cite{goldberg:et:al:21}, which is run
off-chain by an untrusted prover.
Using the STARK cryptographic proof system \cite{ben:sasson:et:al:18},
the prover then publishes a succinct certificate for
the result of the off-chain computation that can be verified efficiently
on chain. The method allows users to trust the claimed result of a computation
without having to carry it out themselves.

More precisely, Cairo is a virtual machine and a family of programming languages
implementing this strategy.
At the most fundamental level, Cairo is a virtual CPU, an instruction set,
and a method of encoding execution traces efficiently with STARK.
At a slightly higher-level of abstraction, programs can be written in the Cairo
assembly language, Casm. At the next level is a programming language,
now called CairoZero, that compiles down to Casm. It adds things like
function declarations, variables, labels, and structured data types.
A newer programming language, now known simply as ``Cairo,'' sits at an even higher level,
offering a more expressive type system and type safety guarantees.
This high-level language is designed to enable users to write and deploy
decentralized applications on Starknet.
CairoZero is still used
for the implementation of Starknet as well as for implementing core library
functions for the new Cairo programming language.
This paper is solely about the verification of programs written in CairoZero;
our verification efforts related to the new Cairo programming language are ongoing and will be described
elsewhere. For the latter, we have ported the Cairo CPU semantics and many of our
libraries to the newest version of the Lean proof assistant, Lean 4 \cite{de:moura:ullrich:21},
but for CairoZero verification we still rely on Lean 3 \cite{de:moura:et:al:15}.

Here we describe an augmentation of the CairoZero compiler that enables users to
produce formal proofs that compiled machine code meets its high-level specifications.
We retain enough information during the compilation phase
for our verification tool to extract a description of the machine code as well as
naive functional specifications of the source code. We automatically construct
formal proofs, in Lean, that the
machine code meets these specifications.
Users can then write their own specifications of the source code in Lean
and prove that they are implied by the automatically generated ones.
In doing so, they can make use of their specifications of functions earlier
in the dependency chain.
Using Lean to check both the user-written and autogenerated proofs yields
end-to-end verification of the user's specifications,
down to CPU semantics. In other work \cite{avigad:et:al:22}, we have moreover
verified the correctness of the algebraic encoding of the CPU semantics that is
used to generate the certificates used in the STARK protocol.

In Sections~\ref{sec:the:cairo:machine:model} to \ref{sec:from:cairo:code:to:assembly}
we describe the Cairo CPU architecture, Casm, and CairoZero,
and we explain how we generate high-level CairoZero specifications and construct formal
correctness proofs in Lean. Section~\ref{sec:memory:management:and:range:checks}
focuses on features of our work that are specific to the domain of application
and the STARK encoding, namely, memory management in CairoZero and mechanisms for
computing with elements of a finite field.
In Section~\ref{sec:elliptic:curve:computations}, we demonstrate the success of our
approach by describing our verification of elliptic curve computations over
large finite fields different from the native field used by the Cairo virtual machine.
Similarly, in Section~\ref{sec:implementing:secp}, we describe our
verification of a procedure for validating digital signatures, and in
Section~\ref{sec:dictionaries}, we describe our verification of a mechanism for
simulating a read-write dictionary data structure in a read-only setting.
In Section~\ref{sec:methodology}, we explain why our approach has been
effective in practice, enabling us to verify production code
without hindering the development of the compiler or the library.
Our main contributions are therefore as follows:
\begin{itemize}
  \item We provide means of obtaining end-to-end verification, in a foundational proof assistant,
  of Cairo machine code with respect to high-level specifications.
  \item We handle novel features of the execution model that are specific to its
  use in blockchain applications.
  \item We demonstrate that our approach scales well by presenting substantial case
  studies, verifying code that is used in production.
  \item We explain how we managed to carry out our work in an industrial setting,
  while the language and compiler were under continuous development.
  \item We explain why our approach, which involves automatically generating
  source-level proofs that are elaborated and checked by Lean,
  has been surprisingly effective.
\end{itemize}
Our Lean libraries, our verification tool, and the case study described here can be found online
at \url{https://github.com/starkware-libs/formal-proofs}.

This a revised and expanded version of a conference paper of the same name \cite{avigad:et:al:23}.
In the conference version, we described the implementation of elliptic curve operations
for the secp256k1 curve
only briefly; since then, we have verified the operations for the secp256r1 curve as well and we describe
both here in greater detail. Our original verification contained one ``sorry,'' which is to say, we assumed the associativity
of the elliptive curve addition law without proof. Section~\ref{sec:elliptic:curve:computations} explains
how we have now eliminated that gap relying on the work of Angdinata and Xu \cite{angdinata:xu:23}.
The verification of the \lstinline{squash_dict} procedure, described in Section~\ref{sec:dictionaries},
is new.
Finally, we have revised and expanded the exposition throughout.

The conference version refers to CairoZero as ``Cairo 0'' or simply ``Cairo'';
we have updated our terminology here.
We still use ``Cairo'' to refer to
the full ecosystem, for example, when we refer to the Cairo virtual machine, Cairo machine code,
and so on. When it is not necessary
to distinguish between them, we use the phrase ``Cairo program'' to stand
generically for a Cairo assembly program,
a CairoZero program, or a program written in the high-level Cairo language.

\section{The Cairo Machine Model}
\label{sec:the:cairo:machine:model}

The Cairo machine model is based on a simple CPU architecture with
three registers: a \emph{program counter} (pc), which points to the current
instruction in memory; a \emph{frame pointer} (fp), which generally points to the
location of the local variables in a function call, and an \emph{allocation pointer} (ap),
which generally points to the next free value in global working memory.
A machine instruction consists of three 16-bit words, which generally
serve as memory offsets for operations performed on memory, and 15 1-bit
flags, which determine the nature of the instruction. The architecture is
described in detail in the Cairo whitepaper \cite{goldberg:et:al:21}, and a Lean formalization thereof
is described in \cite{avigad:et:al:22}.

A notable feature of the machine model is that elements of memory, as well as the
contents of the registers, are elements of the field of integers modulo a certain prime
number (by default, $2^{251} + 17 \cdot 2^{192} + 1$).
This is because execution traces have to be encoded in such a way that their existence can be established
by a STARK crytographic certificate, and, fundamentally, the STARK protocol serves to establish the
existence of solutions to polynomial equations over a finite field \cite{goldberg:et:al:21,avigad:et:al:22}.
The CPU can therefore add and multiply
values, but it cannot ask whether one value is greater
than another. Cryptographic primitives, described in Section~\ref{sec:memory:management:and:range:checks}, can be used to assert that
the contents of a memory location represent the cast of an integer in a certain
range. The core library uses values that are checked to lie in the interval
$[0, 2^{128})$.

Another notable feature of the machine model is that the memory is read-only.
To establish a computational claim on blockchain, a prover makes public a
partial assignment to
memory that typically includes the program that is executed and the agreed-upon
input. The prover also makes public the initial and final state of the registers
and the number of steps in the computation.
A certificate published on blockchain establishes, modulo common cryptographic assumptions,
that the prover is in possession of a full assignment to memory extending the partial one such
that the program runs to completion in the given number of steps. The code is then
carefully designed to ensure that this implies the claim that is of interest
to the verifier.
For example, to establish that a calculation yields a claimed result, the prover
and verifier agree on a Cairo program that carries out the calculation, asserts that it
is equal to the claimed value, and fails otherwise. A certificate
that the program terminates successfully establishes the
computational claim.

Reading Cairo programs---whether written in Cairo assembly, CairoZero, or the new Cairo
language---takes some getting used to. Whereas a program
instruction like {\tt x = y + 5} is often thought of as an assignment of the
value $y + 5$ to the memory location allocated to {\tt x}, in this setting it is really an
assertion that the prover has assigned values to the memory so that the
equation holds. It is an interesting feature of the model that a
Cairo program can depend on values in memory that are
assigned by the prover but not made public to the verifier. For example,
a program can establish that a value {\tt x} is a perfect square by asserting
that it is equal to {\tt y * y} for some {\tt y}, without sharing the value
of {\tt y} with the verifier.

The Cairo CPU instruction set includes a call instruction, a return instruction,
conditional and unconditional jumps, an instruction to advance the allocation pointer,
and instructions that make arithmetic assertions about values stored in memory.
The file {\tt cpu.lean} consists of less than 200 lines of Lean definitions
that provide a formal specification of the CPU and the next state relation.
The next state depends on the contents of memory, \lstinline{mem},
and the values of the CPU registers, \lstinline{s}, but since the memory never
changes, the next state relation \lstinline{next_state mem s t} need only specify the successor
state \lstinline{t} of the
registers. If the program counter points to an assert instruction
that fails, there is no successor state. If the program counter points to an
ill-formed instruction, the value of \lstinline{t} is nondeterministic,
so verifying that a Cairo program has the intended semantics
generally requires establishing that a successful
run of the program does not encounter such an instruction.
By convention, programs halt with a jump instruction to itself,
that is,
an infinite loop. The cryptographic proof published on blockchain
establishes, with high probability, that the program has reached such a state.

More precisely, the Cairo encoding \cite{goldberg:et:al:21} reduces the existence of an
execution trace to the existence of a solution to a system of polynomial equations. The STARK protocol uses a cryptographic scheme \cite{ben:sasson:et:al:18}
to establish, with high probability, that the prover has knowledge of such a solution.
We do not verify the STARK protocol itself, but, in prior work \cite{avigad:et:al:22},
we have verified the encoding scheme for Cairo execution
traces, in other words, the reduction of the execution claim to the polynomial equations.

The project described here picks up where the previous project leaves off:
it is designed to enable users to prove that the successful execution
of a particular program guarantees that a property of interest holds. To that end, we define the
following predicate:
\begin{lstlisting}
def ensures (mem : F → F) (σ : register_state F)
    (P : ℕ → register_state F → Prop) : Prop :=
  ∀ n : ℕ, ∀ exec : fin (n+1) → register_state F,
    is_halting_trace mem exec → exec 0 = σ →
      ∃ i : fin (n+1), ∃ κ ≤ i, P κ (exec i)
\end{lstlisting}
This says that any sequence \lstinline{exec} of states that starts with
\lstinline{σ}, proceeds according to the machine semantics with respect to
the memory assignment \lstinline{mem}, and ends with a halting
instruction eventually reaches a step \lstinline{i} along the way such that the register
state \lstinline{exec i} satisfies \lstinline{P}. Note that the predicate
\lstinline{P} can also make reference to the contents of memory, so we can
express that at step \lstinline{i} the memory location referenced by a certain
register has a certain value, or that the value of a fixed memory location has
a certain property. More precisely, the predicate \lstinline{P κ τ} takes
a numeric value \lstinline{κ} as well as a register state \lstinline{τ}, and the
\lstinline{ensures} predicate says that there is a value of \lstinline{κ} less
than or equal to \lstinline{i} such that \lstinline{P κ (exec i)} holds.
We will explain the use of \lstinline{κ} in Section~\ref{sec:memory:management:and:range:checks}.


\section{From Assembly Code to Machine Code}
\label{sec:from:assembly:code:to:machine:code}

The CairoZero compiler translates code written in the CairoZero language to instructions in the Cairo assembly language \cite[Section 5]{goldberg:et:al:21},
which are then translated to machine instructions. Assembly instructions can also be inserted directly into CairoZero programs.
The first step toward bridging the gap between the CairoZero programming language
and Cairo machine code is therefore to model the Cairo assembly language
 in Lean.
The file {\tt soundness/assembly.lean} in our project provides a description of
Cairo machine instructions in terms of the offsets and flags, and it defines a
translation from that representation to 63-bit machine
code instructions. It also defines Lean notation that
approximates Cairo assembly-language syntax. For example, here are three elementary
mathematical CairoZero functions from the CairoZero common library:
\begin{lstlisting}
func assert_nn{range_check_ptr}(a) {
  a = [range_check_ptr];
  let range_check_ptr = range_check_ptr + 1;
  return ();
}

func assert_le{range_check_ptr}(a, b) {
  assert_nn(b - a);
  return ();
}

func assert_nn_le{range_check_ptr}(a, b) {
  assert_nn(a);
  assert_le(a, b);
  return ();
}
\end{lstlisting}
The first confirms that the argument \lstinline{a} is the cast of a nonnegative
integer less than $2^{128}$, by asserting that it is equal to
the value of memory at the address \lstinline{range_check_ptr},
which is assumed to point to a block of elements that have been verified to
have this property. The second confirms that \lstinline{a} is less than or
equal to \lstinline{b} by calling \lstinline{assert_nn(b - a)}, and the third
function combines the previous two properties.
The curly brackets mean that the argument \lstinline{range_check_ptr} is
passed to, updated by, and returned implicitly by these functions.
We explain range checking in more detail
in Section~\ref{sec:memory:management:and:range:checks}.

The CairoZero compiler compiles these functions to assembly code and then to machine instructions.
The assembly code corresponding to \lstinline{assert_nn_le} looks as follows:
\begin{lstlisting}
[ap] = [fp + (-5)]; ap++
[ap] = [fp + (-4)]; ap++
call rel -11
[ap] = [fp + (-4)]; ap++
[ap] = [fp + (-3)]; ap++
call rel -11
ret
\end{lstlisting}
The details are not important. The function calls are carried out by copying
the arguments, including the implicit range check
pointer, to the end of the global memory used so far. The arguments are referenced relative to the
frame pointer, so \lstinline{[fp + (-5)]} denotes the value in memory at the
address \lstinline{fp - 5}.
The call instructions update the program counter and frame pointer so that execution continues in the subroutine, and the
return at the end restores the frame pointer and updates the program counter to the next instruction in the calling routine.

Our tool generates a Lean description of this assembly code.
The notation isn't pretty; we use tick marks and funny tokens
to avoid conflicting with other tokens that may be in use.
For example, the Lean description of the \lstinline{assert_nn_le} assembly code
is as follows:
\begin{lstlisting}
def starkware.cairo.common.math.code_assert_nn_le : list F := [
  'assert_eq['dst[ap] === 'res['op1[fp+ -5]];ap++].to_nat,
  'assert_eq['dst[ap] === 'res['op1[fp+ -4]];ap++].to_nat,
  'call_rel['op1[imm]].to_nat, -11,
  'assert_eq['dst[ap] === 'res['op1[fp+ -4]];ap++].to_nat,
  'assert_eq['dst[ap] === 'res['op1[fp+ -3]];ap++].to_nat,
  'call_rel['op1[imm]].to_nat, -11,
  'ret[].to_nat ]
\end{lstlisting}
We note in passing that mechanisms for custom syntax in the newer version of Lean, Lean~4,
allow us to use Casm verbatim in a Lean definitions.
But for our work with Lean~3,
we only had to read such code when debugging, and the clunky syntax worked just fine.
Our Lean representation is adequate in that the assembly instructions
can be transformed to machine instructions, which can, in turn, be
transformed to the 63-bit numeric representations which are then
cast to the finite field \lstinline{F}.
Users can evaluate definitions like the one above in Lean and check that the
resulting numeric values are the same ones produced by the CairoZero compiler,
and hence are the same ones used in the STARK certificate generated by the
Cairo runner.
Our soundness proofs start with the assumption that these values are stored
in memory and that the program counter is set accordingly.

The file {\tt soundness/assembly.lean} establishes a small-step semantics
for reasoning about instructions at the assembly level.
For example, variants of the Cairo {\tt call} instruction allow specifying
the address of the target in various ways, either as
an absolute or relative address, which can in turn be given as an immediate
value or read from memory with an offset from either the allocation pointer
or frame pointer. The theorem describing the behavior of this instruction
is as follows:
\begin{lstlisting}
theorem next_state_call {F : Type*} [field F] (mem : F → F)
  (s t : register_state F) (op0 : op0_spec) (res : res_spec)
  (call_abs : bool) :
(call_instr call_abs res).next_state mem s t ↔
  (t.pc = jump_pc s call_abs (compute_res mem s (op0_spec.ap_plus 1) res) ∧
    t.ap = s.ap + 2 ∧
    t.fp = s.ap + 2 ∧
    mem (s.ap + 1) = bump_pc s res.to_op1.op1_imm ∧
    mem s.ap = s.fp)
\end{lstlisting}
Read this as follows: given that the CPU registers are in state {\tt s} and
given the contents of memory {\tt mem}, the call instruction with boolean
flag \lstinline{call_abs} and operand specifications {\tt op0} and {\tt res}
results in the
new state {\tt t}, where the program counter is updated as indicated,
the relevant return address and the current frame pointer are stored in memory
at the current allocation pointer,
the allocation pointer is increased by two, and the frame pointer is increased
by two.
The details of the computations \lstinline{jump_pc}, \lstinline{compute_res},
and \lstinline{bump_pc}
are not important. What \emph{is} important is that for concrete values of
{\tt op0}, {\tt res}, and \lstinline{call_abs}, Lean's tactics
(a term rewriter, a numeric evaluator, etc.) are powerful enough to compute
specific values and \emph{prove} that the functions have those values.
For example, if a specific call instruction decreases the program counter
by 100, Lean can prove that the \lstinline{next_state}
relation holds for a suitable state {\tt t} with {\tt t.pc = s.pc - 100}.
This allows us to reason about the behavior of a block of
assembly code by stepping through each instruction in turn.
The proof of the \lstinline{next_state_call} and others like it are fiddly but straightforward:
it is just a matter of unfolding the definitions of the assembly language
instructions and then relating the resulting machine instructions to the
semantics defined in {\tt cpu.lean}.

\section{From CairoZero Code to Assembly Code}
\label{sec:from:cairo:code:to:assembly}

Consider the procedure \lstinline{assert_nn_le}, which takes field
elements \lstinline{a} and \lstinline{b} and asserts that they are
casts of integers in a certain range such that the one corresponding to
\lstinline{a} is less than or equal to the one corresponding to \lstinline{b}.
More precisely, the desired specification is as follows:
\begin{lstlisting}
def spec_assert_nn_le (mem : F → F) (κ : ℕ)
  (range_check_ptr a b ρ_range_check_ptr : F) : Prop :=
∃ m n : ℕ, m < rc_bound F ∧ n < rc_bound F ∧ a = ↑m ∧ b = ↑(m + n)
\end{lstlisting}
The argument \lstinline{ρ_range_check_ptr} denotes the return value of \lstinline{assert_nn_le},
which is implicit in the CairoZero code. We often use unicode characters, which
are allowed in Lean but not CairoZero, to ensure that identifiers that we introduce
in specifications and proofs do not clash with the identifiers that we take from
CairoZero.
The up arrows denote casts to the field \lstinline{F}.
Here the value of \lstinline{rc_bound F} is assumed to be $2^{128}$, so $m$ and $n$
represent 128-bit unsigned integers.
The autogenerated specification of \lstinline{assert_nn_le} merely says
that the CairoZero code calls the two auxiliary functions \lstinline{assert_nn}
and \lstinline{assert_le}:
\begin{lstlisting}
def auto_spec_assert_nn_le (mem : F → F) (κ : ℕ)
    (range_check_ptr a b ρ_range_check_ptr : F) : Prop :=
∃ (κ₁ : ℕ) (range_check_ptr₁ : F),
  spec_assert_nn mem κ₁ range_check_ptr a range_check_ptr₁ ∧
∃ (κ₂ : ℕ) (range_check_ptr₂ : F),
  spec_assert_le mem κ₂ range_check_ptr₁ a b range_check_ptr₂ ∧
κ₁ + κ₂ + 7 ≤ κ ∧
ρ_range_check_ptr = range_check_ptr₂
\end{lstlisting}
The role of \lstinline{κ}, \lstinline{κ₁}, and \lstinline{κ₂} will be discussed in Section~\ref{sec:memory:management:and:range:checks}.
Notice that the autogenerated specification for \lstinline{assert_nn_le} refers
to the \emph{user} specifications of \lstinline{assert_nn} and \lstinline{assert_le} rather than the autogenerated ones.
Interleaving the two types of specifications is crucial for handling recursion,
since our autogenerated specification of a recursive function invokes the
user specification to describe the effects of the recursive calls.
We handle loops in a similar way.
More importantly, our approach means that when users have to reason about the
autogenerated specification, they can make use of their own specifications of
the dependencies. This enables them to verify complex programs in a modular way.

With the autogenerated specification in hand, the user's task is to write their
own specification of \lstinline{spec_assert_nn_le} and prove that it
follows from the autogenerated one.
Our verification tool then uses that in the proof of the following
theorem, which asserts that the machine code meets the user specification:
\begin{lstlisting}
theorem auto_sound_assert_nn_le
  (range_check_ptr a b : F)
  (h_mem : mem_at mem code_assert_nn_le σ.pc)
  (h_mem_0 : mem_at mem code_assert_nn (σ.pc - 9))
  (h_mem_1 : mem_at mem code_assert_le (σ.pc - 5))
  (hin_range_check_ptr : range_check_ptr = mem (σ.fp - 5))
  (hin_a : a = mem (σ.fp - 4))
  (hin_b : b = mem (σ.fp - 3)) :
ensures mem σ (λ κ τ,
  τ.pc = mem (σ.fp - 1) ∧ τ.fp = mem (σ.fp - 2) ∧ τ.ap = σ.ap + 14 ∧
  ∃ μ ≤ κ,
    rc_ensures mem (rc_bound F) μ (mem (σ.fp - 5)) (mem (τ.ap - 1))
    (spec_assert_nn_le mem κ range_check_ptr a b (mem (τ.ap - 1))))
\end{lstlisting}
The theorem asserts that, given the contents of memory \lstinline{mem} and the
register state \lstinline{σ}, if the code for \lstinline{assert_nn_le} is in
memory at the program counter, the code for the dependencies are in place
as well, and the arguments to the function are stored in memory in the
expected locations indexed by the frame pointer, then any halting computation
eventually returns to the calling function (restoring the program counter
and frame pointer according to the CairoZero language calling conventions) and
ensures that the user specification holds, assuming that certain auxiliary
locations in memory have been range checked (cf.~Section~\ref{sec:memory:management:and:range:checks}).

The proof of \lstinline{auto_sound_assert_nn_le} establishes the correctness
of the autogenerated specification and then applies the user-supplied theorem
that this implies the user's specification.
Generally speaking, the user doesn't need to see the Lean description of the
assembly code, the theorem \lstinline{auto_sound_assert_nn_le}, or the proof of correctness.
The autogenerated specifications, the user specifications, and the proof
that the former imply the latter are stored in the same directory as the CairoZero
code. The Lean descriptions of the assembly code and
the correctness proofs are kept in a separate folder, tucked out of
sight.

Our verification tool has the task of extracting the autogenerated
specifications and constructing the correctness proofs.
The CairoZero compiler, which is written in Python,
produces a number of data structures that we are able to make use of once
the compilation is complete.
These contain, for example, a dictionary of namespaced identifiers.
Whenever we needed additional information, we added hooks that, in verification
mode, are called to log that information.
For example, a compound assertion like \lstinline{x = 3 * y + 4 * z} translates
to a list of atomic assertions, and our verification tool has access to
the original equation
and the code points that mark the beginning and end of the list of assertions.

As we will discuss in Section~\ref{sec:memory:management:and:range:checks}, the
CairoZero language uses two sorts of variables: local variables are indexed
with an offset from the frame pointer and global variables are indexed with an
offset from the allocation pointer.
When a function takes values \lstinline{a b c : F}
as arguments, the compiler places these values in memory
just before the allocation pointer when the procedure is called.
For the most part, the verifier keeps the nitty-gritty memory allocation issues
hidden from the
user, and mediates between variable names and the machine semantics with
equations like \lstinline{a = mem (σ.ap - 3)}.
The CairoZero language also allows the
definition of compound structures, and we define the corresponding structures in Lean and interpret references to memory accordingly.

Our verifier uses Dijkstra's weakest preconditions \cite{dijkstra:75}
to read off a specification. The process is straightforward, modulo the fact
that the verifier also has to construct Lean proofs that prove that
these specifications are met.
That requires unpacking the meaning of each machine instruction, using the
theorems described in the previous section. Unpacking the \lstinline{mem_at} predicate
tells us which instruction is present at each memory location. We then use
special-purpose tactics (small-scale automation written in Lean) to unpack
the effect of each instruction.

For example, suppose at a given point in a
proof we need to show that executing the code at program counter
\lstinline{σ.pc + 5} ensures that a certain result holds, and we know
that the instruction at that location corresponds to an \lstinline{assert},
requiring
that the memory \lstinline{mem (σ.ap + 2)} at location
\lstinline{σ.ap + 2} is equal to the immediate argument in the memory at
\lstinline{σ.pc + 6}. Moreover, suppose that one of the \lstinline{mem_at} assumptions entails that this immediate value is equal to 5.
Applying the relevant tactic to step through the code tells us that it suffices
to show that the result of executing the code starting at the next
instruction, at counter \lstinline{σ.pc + 7},
ensures that the desired result holds, under the additional assumption
that the value in memory at \lstinline{σ.ap + 2} is equal to 5.
This reduction is justified
by appealing to the meaning of the \lstinline{ensures} predicate and the specification
of the machine semantics. In this way, our tactics carry out a kind of symbolic
execution of the assembly code, and register the effects in the proof context.

The verifier's task is then to parse each high-level CairoZero instruction, generate a specification of its behavior, and construct the corresponding part
of the correctness proof, which shows that the corresponding assembly instructions
implement the high-level ones.

\begin{itemize}
\item A variable declaration like \lstinline{tempvar b = a + 5} translates to
an existential quantifier in the specification, \lstinline{∃ b, b = a + 5 ∧ ...}.
The correctness proof instantiates the existential quantifier to the corresponding
memory location and maintains this correspondence.

\item An equality assertion in the program translates to an equality assertion in the
specification. Such an assertion generally translates to one or more
assembly-level assertions, and the correctness proof involves reconstructing the compound
equality statement from the components.

\item CairoZero programs can have labels and both conditional and unconditional jumps.
The corresponding machine code has the expected effect of (conditionally)
modifying the program counter. In some settings, the correctness proof
only needs to record this change to the program counter and continue stepping through
the instructions starting at the new pc. But for handling jumps that coalesce
control flow, and loops in particular, we analyze the control flow into blocks
and break the specification and correctness proof up accordingly. We describe
this process further below.

\item A conditional jump is implicit in a CairoZero
\lstinline{if ... then ... else} construct. This
translates to a disjunction in the specification and the definition of a block
where the branches flow together.

\item A subroutine call to another procedure translates to an assertion,
in the autogenerated specification, that the user specification of the
target procedure holds of the arguments and the return value. The call instruction
stores the current program pointer and frame pointer in memory and jumps
to the location of the target procedure. The return instruction restores
the frame pointer and jumps back to the calling procedure.
The correctness proof invokes the
correctness theorem for the target procedure as well as the assumption that
the procedure is in memory at the expected location.
\end{itemize}

The description so far presupposes that the control flow has no cycles. To handle recursive calls and loops, we do not have to prove termination; the STARK certificate assures a skeptical verifier
that the program has terminated, so we need only
show that, given that fact, the specification is met. (This
is commonly characterized as the difference between \emph{partial correctness}
and \emph{total correctness}.) The claim \lstinline{ensures mem σ P} is
equivalent to \lstinline{∀ b, ensuresb b mem σ P},
where \lstinline{ensuresb b mem σ P} says that every halting execution sequence
from state \lstinline{σ} with at most \lstinline{b} steps eventually reaches a
state that satisfies \lstinline{P}. We can prove the latter by induction on
\lstinline{b}, generalizing over states \lstinline{σ} with program counter
pointing to the relevant code.

For functions that call themselves recursively, we modify the default user
specification so that it is trivially true, and place it \emph{before} the
autogenerated specification. The autogenerated specification
asserts the play-by-play description alluded to above, except that it uses the
user specification to characterize the recursive calls. The user is free to
write any specification they want, provided they show that the autogenerated
specification implies the user specification. In short, the user has to show
that the play-by-play characterization implies their own characterization,
assuming their characterization holds at downstream calls. The correctness
proof uses this together with the inductive hypothesis at downstream calls.

A similar method handles loops. Our verification tool begins by analyzing
the control-flow graph \cite{allen:70} and dividing the code into basic blocks,
which do not have any jumps or labels.
A block starts at the beginning of a function or at a label,
and ends with a jump, a return, or a flow to a label. Any block that can be
entered from more than one other block receives a separate specification in
the specification file, and cycles that arise in a topological sort are
handled in a manner similar to recursive function calls. In practice,
the user can specify that the execution has an effect conditional on
an invariant holding at the entry point. Verification of
the full user specification then requires showing that the
invariant holds at the first entry point and that it is maintained
on re-entry.

\section{Memory Management and Range Checks}
\label{sec:memory:management:and:range:checks}

In this section, we discuss aspects of the CairoZero programming language that stem
specifically from its intended use toward verifying computations on blockchain.
Encoding execution traces efficiently required keeping the machine model simple,
which is why StarkWare's engineers settled on a CPU with only three registers and
read-only memory. The cost of certification on blockchain scales with the number
of steps in the execution trace together with the number of memory accesses.
The fact that memory is read-only makes the verification task easier: we do
not need to worry about processes overwriting each other's memory.
The CairoZero language allows procedures to declare two types of temporary variables,
namely, relative to the frame pointer (fp) or allocation pointer (ap). It is a quirk of the
CairoZero language that references to ap-based variables can be revoked when the
compiler cannot reliably track the effects of intermediate commands and function
calls on the ap, for example, when different flows of control may result in
different changes to the ap. Our verification relies on the compiler's internal
record of its ability to track these changes.

The STARK encoding is most efficient when memory is assigned in a continuous block.
The CairoZero compiler is tightly coupled with a Cairo \emph{runner},
whose task is to allocate memory and assign values to ensure that the CairoZero program
runs to completion. The Cairo whitepaper describes methods that are used
to simulate conventional memory models, such as the method of implementing
read-write dictionaries that is described in Section~\ref{sec:dictionaries}.
The processor uses the frame pointer to
point to the base of a procedure's local memory and the allocation pointer to point to
the next available position in global memory. Local variables are kept in the
same contiguous block.
When one procedure calls another, the program counter and frame pointer
are stored in global memory, the allocation pointer is updated, and the frame
pointer is set equal to the allocation pointer. When the procedure returns,
the program counter and frame pointer are restored.

For efficiency, a procedure sometimes has to access values that are
stored in global memory, which is to say, they are indexed relative to the
allocation pointer. This is challenging because the allocation pointer is
constantly changing. For example, when one procedure calls another, upon
the return the allocation pointer may have changed. Moreover, the new value
of the allocation pointer cannot always be predicted at compile time; for example,
different flows of control through an if-then-else can result in different
changes to its value, and the change incurred by a call to a recursive function
will generally depend on its runtime arguments.
The compiler uses a \emph{flow tracker} that keeps
track of these changes as best it can, allowing the programmer to refer to
the same global variables throughout.
When it can't determine the reference of an ap-based variable, it raises an error
that tells the programmer that the reference has been ``revoked.''
This requires the programmer to either use an fp-based variable or find another workaround.
Our verification tool does not have
to know much about how the flow tracker works, but it needs to make use of
the results. For example, if the value of a variable \lstinline{x} is
\lstinline{mem (ap + 1)} before a subroutine call and the allocation pointer
\lstinline{ap'} on return is equal to \lstinline{ap + 3}, our Lean proofs
need to use the identity \lstinline{ap' = ap + 3} to translate
an assertion involving {\lstinline{mem (ap' - 2)}} into an assertion about
\lstinline{x}. Our verification tool claims and proves the relevant identities while
stepping through the code,
and uses those identities as rewriting rules when verifying assertions.

A more striking difference between programming in CairoZero and programming in an
ordinary programming language is that the most fundamental data type consists
of values of a finite field. One can add and multiply field elements, but there
is no machine instruction that compares the order of two elements.
To meet high-level specifications that are stated in
terms of integers, the STARK encoding uses cryptographic primitives to
verify that a specified range of memory has been \emph{range checked}, which
is to say, the corresponding field elements are casts of integers in the
interval $[0, 2^{128})$. Our previous verification of the STARK encoding
\cite{avigad:et:al:22}
shows that the STARK certificate guarantees
(with high probability; cf.~\cite[Sections 9 and 10]{avigad:et:al:22}) that
the specified memory locations have indeed been range checked. A CairoZero program
can make use of this fact by taking, as input, a pointer to a location in
the block of range-checked memory, making assertions about a sequence of
values at that location, and returning (in addition to its ordinary return
values) an updated pointer to the next unused element. Both the user specification
and the autogenerated specification are of the form ``assuming the values
between \ldots{} and \ldots{} have been range-checked, the following holds: \ldots\,.''
These hypotheses have to be used inside the correctness proofs, to justify
the assertions that particular values have been range checked. The hypotheses
also have to be threaded through procedure calls and combined appropriately,
so the specification of a top-level function comes with a range-check hypothesis
that covers all the recursive calls. This top-level hypothesis is
justified by the STARK certificate.

Our verification tool handles all this plumbing. For example, recall the CairoZero
function \lstinline{assert_nn}, which asserts that the argument
\lstinline{a} is the cast of an integer in $[0, 2^{128})$.
\begin{lstlisting}
func assert_nn{range_check_ptr}(a) {
  a = [range_check_ptr];
  let range_check_ptr = range_check_ptr + 1;
  return ();
}
\end{lstlisting}
The curly brackets in \verb~{range_check_ptr}~ indicate that the value
should implicitly be returned among the other return values. (In this case,
there aren't any others.) The user-written specification of this function is as follows:
\begin{lstlisting}
def spec_assert_nn (mem : F → F) (κ : ℕ)
    (range_check_ptr a ρ_range_check_ptr : F) : Prop :=
  ∃ n : ℕ, n < rc_bound F ∧ a = ↑n
\end{lstlisting}
Recall that the annotation \lstinline{↑n} casts the natural number \lstinline{n}
to the underlying field \lstinline{F}.
Our verification tool generates the following specification:
\begin{lstlisting}
def auto_spec_assert_nn (mem : F → F) (κ : ℕ)
    (range_check_ptr a ρ_range_check_ptr : F) : Prop :=
  a = mem (range_check_ptr) ∧
  is_range_checked (rc_bound F) a ∧
  ∃ range_check_ptr₁ : F, range_check_ptr₁ = range_check_ptr + 1 ∧
  3 ≤ κ ∧
  ρ_range_check_ptr = range_check_ptr₁
\end{lstlisting}
Here, \lstinline{range_check_ptr₁} is the updated version of \lstinline{range_check_ptr},
and the last line specifies that this is the function's sole return value.
Our tool detects the reference to \lstinline{range_check_ptr} in the CairoZero code and
adds \lstinline{is_range_checked (rc_bound F) a} to the autogenerated specification,
generating an obligation in the correctness proof that the user does not have to see.
The user has to prove that \lstinline{spec_assert_nn} follows from
\lstinline{auto_spec_assert_nn}, but this follows immediately from the conjunct
\lstinline{is_range_checked (rc_bound F) a} in the autogenerated specification.
Notice that \lstinline{auto_spec_assert_nn} asserts
\lstinline{3 ≤ κ}, which signifies that the execution of the function adds at least
three steps to the trace.
This fact is discarded in the proof of \lstinline{spec_assert_nn}, but
for an example of a specification that uses \lstinline{κ}, see Section~\ref{sec:dictionaries}.

The ultimate correctness theorem is stated as follows:
\begin{lstlisting}
theorem auto_sound_assert_nn
  (range_check_ptr a : F)
  (h_mem : mem_at mem code_assert_nn σ.pc)
  (hin_range_check_ptr : range_check_ptr = mem (σ.fp - 4))
  (hin_a : a = mem (σ.fp - 3)) :
ensures mem σ (λ κ τ,
  τ.pc = mem (σ.fp - 1) ∧ τ.fp = mem (σ.fp - 2) ∧ τ.ap = σ.ap + 1 ∧
  ∃ μ ≤ κ, rc_ensures mem (rc_bound F) μ (mem (σ.fp - 4)) (mem (τ.ap - 1))
    (spec_assert_nn mem κ range_check_ptr a (mem (τ.ap - 1))))
\end{lstlisting}
In this specification, the assertions \lstinline{τ.pc = mem (σ.fp - 1)}
and \lstinline{τ.fp = mem (σ.fp - 2)} assert that the program counter and
frame pointer have been restored correctly when the function returns.
Our verification tool learns from the flow tracker that
any path through this code updates the allocation pointer by one, and so
it also establishes that fact, i.e.~\lstinline{τ.ap = σ.ap + 1}, to
make that information accessible when reasoning about procedures that call it.
The \lstinline{rc_ensures} clause in the conclusion says that
if the block of memory between
\lstinline{mem (σ.fp - 4)} and \lstinline{mem (τ.ap - 1)} is range-checked
then the user specification holds.
(We will return to the role of \lstinline{μ} in a moment.) Here \lstinline{σ.fp}
is the value of the frame pointer when the function is called, \lstinline{τ.ap}
refers to the value of the allocation pointer upon return, and
\lstinline{mem (σ.fp - 4)} and \lstinline{mem (τ.ap - 1)} are, respectively, the location of
the argument \lstinline{range_check_ptr} and the return value, which is
supposed to be the updated range check pointer.

We can now explain the role of \lstinline{κ} and \lstinline{μ}.
Recall that \lstinline{mem (σ.fp - 4)} and \lstinline{mem (τ.ap - 1)} are
field elements. A first guess as to how to specify that the range of
memory values between those two locations is range checked is to
say that there is a natural number \lstinline{μ} such that
\lstinline{mem (τ.ap - 1) = mem (σ.fp - 4) + ↑μ} and for every \lstinline{i < μ},
the value in memory at address \lstinline{mem (τ.ap - 1) + i} is range checked.
But this specification is problematic: if the equation holds for some small
value of \lstinline{μ}, it also holds for \lstinline{μ} plus the characteristic
of the underlying field, since the characteristic is cast to 0.
Therefore, the mere existence of such a \lstinline{μ} does not imply the soundness
claim, and our correctness proof needs to make use of the fact that the
total number of range-checked elements \lstinline{μ} does not wrap around
the finite field.
We achieve this by asserting that \lstinline{μ} is, moreover, bounded by
the number of steps \lstinline{κ} in the execution trace, which is made public
in the STARK
certification and is always smaller than the characteristic of the field.
In the case of range checks, the bounds are handled entirely by the verifier
and the user need not worry about them. But we have found that some CairoZero
specifications require similar reasoning about bounds on the length of the
execution, and for those rare occasions, we have exposed the parameter
\lstinline{κ} in the user-facing specifications. We describe one such
use of \lstinline{κ} in Section~\ref{sec:dictionaries}.

The virtue of our verification tool is that the user can be oblivious
to most of the implementation details we have just described, such as the
handling of the range check pointers and the way that variables,
arguments, and return values are stored in memory. The user writes
the CairoZero procedure \lstinline{assert_nn} and is given the specification
\lstinline{auto_spec_assert_nn}.
The user then writes the specification
\lstinline{spec_assert_nn} and proves that \lstinline{spec_assert_nn} follows
from \lstinline{auto_spec_assert_nn}.
The correctness proof can be checked behind the scenes.
From that moment on, \lstinline{spec_assert_nn} is all that users need to know
about the behavior of \lstinline{assert_nn},
from the point of view of proving properties of CairoZero functions that use it.
In the next three sections, we will show that this scales
to the verification of more complex programs.

\section{Elliptic Curve Computations}
\label{sec:elliptic:curve:computations}

Any elliptic curve over a field of characteristic not equal to 2 or 3 can
be described as the set of solutions to an equation $y^2 = x^3 + a x + b$,
the so-called \emph{affine} points, together with one additional \emph{point
at infinity}. Assuming the curve is nonsingular, the set of such points has the
structure of an abelian group,
where the zero is defined to be the point at infinity and addition between affine points
defined as follows:
\begin{itemize}
  \item To add $(x, y)$ to itself, let $s = (3x^2 + a) / 2y$, let $x' = s^2 - 2x$,
    and let $y' = s (x - x') - y$. Then $(x, y) + (x, y) = (x', y')$. This is
    known as \emph{point doubling}.
  \item $(x, y) + (x, -y) = 0$, that is, the point at infinity. In other words,
    $-(x, y) = (x, -y)$.
  \item Otherwise, to add $(x_0, y_0)$ and $(x_1, y_1)$, let
    $s = (y_0 - y_1) / (x_0 - x_1)$, let $x' = s^2 - x_0 - x_1$, and let
    $y' = s (x_0 - x') - y_0$. Then $(x_0, y_0) + (x_1, y_1) = (x', y')$.
\end{itemize}
It is not hard to prove that with addition, negation, and zero so defined, the
structure satisfies all the axioms for an abelian group other than
associativity. Proving associativity is trickier, though it can be done with
brute-force algebraic computations in computer algebra systems, and various
approaches have been used in the interactive theorem proving literature to
establish the result formally
\cite{thery:07,bartzia:strub:14,erbsen:17,hales:raya:20,angdinata:xu:23}.

The study of elliptic curves over the complex numbers, where the addition law has a geometric interpretation, originated in the
nineteenth century.
The topic is fundamental to contemporary number theory.
Elliptic curves over a
finite field are widely used in cryptography today, on the grounds that
for any nonzero point $x$, the map $n \mapsto n \cdot x$ (that is, $n$-fold sum
of $x$ with itself) is easy to compute but, as far as we know, difficult to
invert. This forms the basis for the \emph{elliptic curve digital signature
algorithm} (ECDSA), which we describe in the next section.

The CairoZero library contains functions that compute the scalar product $n \cdot x$
of a natural number $n$ and an element $x$ of either the secp256k1 or secp256r1
elliptic curve. The first of these is the curve $y^2 = x^3 + 7$ over the finite
field of integers modulo the prime $p = 2^{256} - 2^{32} - 977$,
whereas the second is the curve $y^2 = x^3 - 3 x + b$ over the finite
field of integers modulo the prime $p = 2^{256} - 2^{224} + 2^{192} + 2^{96} - 1$,
where $b$ is a certain large number.\footnote{In hexadecimal, $b$ is 0x5ac635d8aa3a93e7b3ebbd55769886bc651d06b0cc53b0f63bce3c3e27d2604b.}
For reasons we will shortly explain, the calculations are subtle, and so an
important target for verification.

In each case, we define the relevant elliptic curve in a file,
\verb|elliptic_curves.lean|, over an arbitrary field $F$ of
characteristic not equal to 2. A point is either the zero point or an \emph{affine point}
given by a pair $(x, y)$ satisfying the curve equation.
\begin{lstlisting}
structure AffineECPoint (F : Type) [field F] :=
(x y : F) (h : on_ec (x, y))

inductive ECPoint  (F : Type) [field F]
| ZeroPoint : ECPoint
| AffinePoint : AffineECPoint F → ECPoint
\end{lstlisting}
We define the group operations described above and prove they are
well defined, for example, that the sum of two points on the curve is
again on the curve. These definitions are needed to state the
specifications of the CairoZero functions we verified. We also prove that the
group operations form a group. Previously, we
omitted only a proof of associativity, using the \lstinline{sorry} keyword
in Lean as a placeholder. As we were preparing the secp256k1 verification
for publication
\cite{avigad:et:al:23}, David Angdinata and Junyan Xu proved, in Lean,
that the elliptic curve law forms a group, in
impressive generality \cite{angdinata:xu:23}. This has since enabled us to eliminate
the \lstinline{sorry} by showing that our representations are isomorphic
to theirs, requiring only about 50 lines of code. Even if the work of
Angdinata and Xu had been available to us from the start, we would likely have
chosen to define something like our current representations
to verify the explicit CairoZero calculations, rather than
unwrap the more abstract definitions in every proof to get at the concrete calculations.

In the CairoZero source, the files \verb|constants.cairo|, \verb|bigint.cairo|,
\verb|field.cairo|, and \verb|ec.cairo| implement the operations over
the secp curve, culminating in an efficient procedure to carry out scalar
multiplication.
The main reason that the code is subtle is that it requires calculations in the
field $\mathbb{Z} / p \mathbb{Z}$, where $p$ is the relevant secp prime,
which is even
larger than (and different from!) the characteristic of the field that underlies
the Cairo machine model.
The CairoZero implementation
thus represents a value $x$ in $\mathbb{Z} / p \mathbb{Z}$ by three field
elements, each of which is checked to be the cast of an integer in a certain range.
We impose additional bounds and hypotheses on these representations, and ensure
that they are maintained by the calculations.
The verification of the secp256k1 and secp256r1 operations
are similar, but in each case the code is carefully adapted to specific features
of the computation.
We will focus on the latter, because the bounds we had to
contend with are even tighter there and because the implementation of
scalar multiplication includes an additional optimization,
which we describe below.

In greater detail, the CairoZero code defines a constant {\tt BASE}, equal to $2^{86}$,
and a structure \verb|BigInt3 {d0: felt, d1: felt, d2: felt}|,
where the \lstinline{felt}
data type refers to an element of the field $F$ underlying the Cairo machine
model. The intention is that the field elements \lstinline{d0}, \lstinline{d1},
and \lstinline{d2} will always be casts of integers \lstinline{i0},
\lstinline{i1}, and \lstinline{i2}, respectively, with absolute values bounded
by something considerably smaller than the characteristic of $F$.
These are intended to represent the value
$\mathtt{i0} + \mathtt{i1} \cdot \mathtt{BASE} + \mathtt{i2} \cdot \mathtt{BASE}^2$,
or, more precisely, the value in $\mathbb{Z} / p \mathbb{Z}$, where $p$ is the
prime characteristic of the secp field.
Our Lean verification has to mediate between at least three different
representations:
\begin{itemize}
  \item Elements $x$ of the secp field of integers modulo the secp prime number.
  \item Triples $(i_0, i_1, i_2)$ of integers, suitably bounded, that represent
    such elements.
  \item Triples of elements $(d_0, d_1, d_2)$ of the underlying field
    \lstinline{F} of the Cairo machine model, assumed or checked to be
    casts of such integers.
\end{itemize}
Field operations like addition and multiplication on the secp field correspond
to addition and multiplication on the integer representations modulo
the secp prime. These in turn are carried out by CairoZero code on the triples
of field elements, with care to ensure that the results track the corresponding
operations on the integer representations.
CairoZero functions dealing with elements of the secp field represented by
elements of type \lstinline{BigInt3} generally assume, as preconditions,
that the elements $d_0, d_1, d_2$ are casts of integers in a certain range
(which varies from function to function), and they generally use range checks
to ensure that their results are similarly bounded.
In the CairoZero code, those preconditions and postconditions were expressed
in comments. A large part of our verification tasks was to ensure that these
comments are correct, in other words, to show that the preconditions are sufficient to
imply the postconditions and that the contracts compose as claimed.

For example, a CairoZero procedure for adding two secp field elements with
different $x$ coordinates has the following signature:
\begin{lstlisting}
func fast_ec_add{range_check_ptr}(point0: EcPoint, point1: EcPoint) -> (res: EcPoint)
\end{lstlisting}
Here, the structure \lstinline{EcPoint} is defined as a pair of \lstinline{BigInt3}
values:
\begin{lstlisting}
struct EcPoint {
  x: BigInt3,
  y: BigInt3,
}
\end{lstlisting}
The preconditions, however, assume that the \lstinline{BigInt3} values
are casts of suitably bounded integers that represent points on the elliptic
curve. We therefore define a structure \lstinline{bigint3}
representing triples of integers and another structure
\lstinline{BddECPointData secpF pt} that contains such triples of integers,
\lstinline{ix} and \lstinline{iy}, satisfying the relevant properties:
\begin{lstlisting}
structure BddECPointData {mem : F → F} (secpF : Type*) [field secpF]
  (pt : EcPoint mem) :=
(ix iy : bigint3)
(ixbdd : ix.bounded' D2_BOUND)
(iybdd : iy.bounded' D2_BOUND)
(ptxeq : pt.x = ix.toBigInt3)
(ptyeq : pt.y = iy.toBigInt3)
(onEC : pt.x = ⟨0, 0, 0⟩ ∨
  (iy.val : secpF)^2 = (ix.val : secpF)^3 + ALPHA * ix.val + BETA)
\end{lstlisting}
Here \lstinline{ixbdd} and \lstinline{iybdd} are the boundedness claims,
\lstinline{ptxeq} and \lstinline{ptyeq} assert that \lstinline{ix} and
\lstinline{iy} represent \lstinline{pt.x} and \lstinline{pt.y}, respectively,
in the Cairo field, and \lstinline{onEC} says that either \lstinline{pt} is the
point at infinity (represented by the CairoZero code by setting \lstinline{pt.x} to
a triple of zeros) or that the pair of values in the secp field represented by \lstinline{(ix, iy)} satisfy the curve equation.
With all this in hand, we can write the specification for \lstinline{fast_ec_add}:
\begin{lstlisting}
def spec_fast_ec_add {mem : F → F} (pt0 pt1 : EcPoint mem)
  (ret1 : EcPoint mem) (secpF : Type) [secp_field secpF] : Prop :=
∀ h0 : BddECPointData secpF pt0,
∀ h1 : BddECPointData secpF pt1,
  ((↑(h0.ix.val) : secpF) ≠ ↑(h1.ix.val) ∨ h0.toECPoint = 0 ∨
    h1.toECPoint = 0) →
  ∃ hret : BddECPointData secpF ret1,
    hret.toECPoint = h0.toECPoint + h1.toECPoint
\end{lstlisting}
In words: assuming \lstinline{pt0} and \lstinline{pt1} are points on the elliptic
curve equipped with \lstinline{BddECPointData} representations \lstinline{h0}
and \lstinline{h1}, and assuming the $x$-coordinates are different or that
both points are the zero point, the function's output value, \lstinline{ret1},
is also equipped with a \lstinline{BddECPointData} representation, and,
when viewed as an element of the secp group, is equal to the sum of the input
points. A more general but less efficient variant of this function, \lstinline{ec_add}, removes
the restriction that the $x$-coordinates are different, which is to say,
it handles $e + e$ and $e + -e$ as well.

For both the secp256k1 and secp256r1 elliptic curves, scalar multiplication
is implemented by iterated doubling and addition.
For example, for the secp256r1 curve, the CairoZero function \lstinline{ec_mul_by_uint256}
takes a point on the elliptic curve and a 256-bit integer (represented by
a pair of field elements that are casts of 128-bit integers) and returns
the scalar product:
\begin{lstlisting}
func ec_mul_by_uint256{range_check_ptr}(point: EcPoint, scalar: Uint256) ->
  (res: EcPoint)
\end{lstlisting}
For efficiency, it uses \lstinline{fast_ec_mul} instead of \lstinline{ec_mul},
which means that, along with the various loop invariants that need to be maintained,
it needs to ensure that the preconditions of \lstinline{fast_ec_mul} are met.
The code also contains an optimization, namely, the first sixteen multiples
of the input point are precomputed and stored in a table for use in the
computation. This makes the invariants even more subtle and difficult to get
right, which, in turn, makes verification even more pressing.
The final specification is as follows:
\begin{lstlisting}
def spec_ec_mul_by_uint256 (mem : F → F) (κ : ℕ) (range_check_ptr : F)
    (point : EcPoint mem) (scalar : Uint256 mem) (ρ_range_check_ptr : F)
    (ρ_res : EcPoint mem) : Prop :=
  ∀ hpt : BddECPointData secpF point,
    ∃ n0 < 2^128, scalar.low = ↑n0 ∧
    ∃ n1 < 2^128, scalar.high = ↑n1 ∧
    ∃ hres : BddECPointData secpF ρ_res,
      hres.toECPoint = (2^128 * n1 + n0) • hpt.toECPoint
\end{lstlisting}
In words, if \lstinline{point} is a point on the elliptic curve
represented by the data \lstinline{hpt}, then the input \lstinline{scalar} represents
a pair of 128-bit unsigned integers $n_1$ and $n_2$, and the output,
represented by \lstinline{hres}, is $2^{128} n_1 + n_0$ times the input.
Note that this procedure explicitly checks that \lstinline{scalar} represents
a 256-bit unsigned integer rather than assumes it.

\section{Validating Digital Signatures}
\label{sec:implementing:secp}

The elliptic curve digital signature algorithm, ECDSA,
provides a protocol by which a sender can generate
a pair consisting of a public key and a private key, publish the public key,
and then send messages in such a way that a receiver can verify that the message
was sent by the holder of the private key and that the message has not been
changed.
We have proved the correctness of a CairoZero procedure for validating cryptographic
signatures using the secp256k1 elliptic curve.
The CairoZero procedure for recovering the public key of the sender
from a digitally signed message looks as follows:
\begin{lstlisting}
func recover_public_key{range_check_ptr}(msg_hash: BigInt3,
  r: BigInt3, s: BigInt3, v: felt) -> (public_key_point: EcPoint) {
alloc_locals;
let (local r_point: EcPoint) = get_point_from_x(x=r, v=v);
let (generator_point: EcPoint) = get_generator_point();
let (u1: BigInt3) = div_mod_n(msg_hash, r);
let (u2: BigInt3) = div_mod_n(s, r);
let (point1) = ec_mul(generator_point, u1);
let (minus_point1) = ec_negate(point1);
let (point2) = ec_mul(r_point, u2);
let (public_key_point) = ec_add(minus_point1, point2);
return (public_key_point); }
\end{lstlisting}
The goal of this section is to explain what this procedure does and the
specification that we verified.

The digital signature method used by Cairo requires
that the sender and receiver agree on the elliptic curve they are using
(in our case, secp256k1) and
on a message hash function. They also fix a point $G$ on the curve that
generates the group, which has a known prime order $n$.
The sender applies a hash function to the message to obtain an integer $m$,
and the method provides a recipe for the sender to generate a triple
$(r, s, v)$ where $r$ and $s$ are integers and $v$ is an additional bit.
The recipient of the message and the signature applies the hash function to
obtain $m$ as well, checks to make sure $r \ne 0$, then finds a point
$(r, y)$ on the elliptic curve by
finding the residues $y$ satisfying $y^2 = r^3 + 7$ in the secp field and
choosing the one that has the same parity as $v$. The receiver then computes
$u_1 = r^{-1} \cdot m$ and $u_2 = r^{-1} \cdot s$, where these operations
take place in the group of residues modulo $n$. Using scalar multiplication,
the receiver calculates $Q = -u_1 \cdot G + u_2 \cdot (r, y)$, a point on the
elliptic curve. If the value $Q$ matches the sender's public key, the receiver
has the desired confirmation that the message $m$ has been sent by the sender.

The procedure \lstinline{recover_public_key} carries out exactly the calculation
of $Q$. It takes as input elements \lstinline{msg_hash}, \lstinline{r},
\lstinline{s}, and \lstinline{v} in the Cairo field.
In the specification,
the first three are assumed to be casts of suitably bounded integers
\lstinline{imsg}, \lstinline{ir}, and \lstinline{is}. In other words, these
assumptions should be guaranteed by the calling procedure.
The specification then asserts that \lstinline{v} is the cast of a suitably bounded
natural
number \lstinline{nv} (this representation is necessarily unique),
and it confirms the existence of data \lstinline{hpoint} representing
the point $-u_1 \cdot G + u_2 \cdot (r, y)$ in the calculation above.
\begin{lstlisting}
def spec_recover_public_key (mem : F → F) (κ : ℕ)
  (range_check_ptr : F) (msg_hash r s : BigInt3 F)
  (v ρ_range_check_ptr : F) (ρ_public_key_point : EcPoint F) : Prop :=
∀ (secpF : Type) [secp_field secpF], by exactI
r ≠ ⟨0, 0, 0⟩ →
∀ ir : bigint3, ir.bounded (3 * BASE - 1) → r = ir.toBigInt3 →
∀ is : bigint3, is.bounded (3 * BASE - 1) → s = is.toBigInt3 →
∀ imsg : bigint3, imsg.bounded (3 * BASE - 1) →
    msg_hash = imsg.toBigInt3 →
∃ nv : ℕ, nv < rc_bound F ∧ v = ↑nv ∧
∃ iu1 iu2 : ℤ,
  iu1 * ir.val ≡ imsg.val [ZMOD secp_n] ∧
  iu2 * ir.val ≡ is.val [ZMOD secp_n] ∧
∃ ny : ℕ, ny < SECP_PRIME ∧ nv ≡ ny [MOD 2] ∧
∃ h_on_ec : @on_ec secpF _ (ir.val, ny),
∃ hpoint  : BddECPointData secpF ρ_public_key_point,
  hpoint.toECPoint =
    -(iu1 • (gen_point_data F secpF).toECPoint) +
      iu2 • ECPoint.AffinePoint ⟨ir.val, ny, h_on_ec⟩
  \end{lstlisting}
The implementation of the procedure requires subtle calculations and checks.
The best explanation of what the intermediate calculations are supposed to
achieve are given by the Lean specification files themselves.

To convey a sense of the effort involved in verifying the secp256k1 elliptic
curve operations and signature validation we can provide some data.
The formalization uses the library file \lstinline{math.cairo}, which runs
about 450 lines of code; our Lean specifications of those functions, as well as our proofs of our own
specifications from the autogenerated ones, comprises about 1,150 lines of code.
The elliptic curve operations and the signature validation procedure run
about 800 lines of CairoZero code and our specification
files run about 3,200 lines of Lean code, on top of about 150 lines in our
definition of the elliptic curve group. The dependency chain of
\lstinline{recover_public_key} consists of 24 CairoZero functions, which compile
to about 900 lines of assembly code, i.e.~900 field elements.
Our autogenerated correctness proofs run about 7,500 lines of Lean code.

Verifying an early version of the CairoZero code turned up two errors that were
independently caught and fixed by the software engineers. The verification later turned up an
error that they missed, having to do with the use of the parameter
\lstinline{v} in \lstinline{recover_public_key}. The error, which does not
allow the prover to fake a signature but does allow it to claim that a valid
signature is invalid, was fixed in the next CairoZero release. Beyond that,
the verification provided the software engineers with welcome reassurance.
Despite extensive code review, they recognized that there were
a number of places where small errors may have crept into the code, and
they were able to breathe a sigh of relief when the verification was complete.

\section{Dictionaries in a Read-Only Memory Model}
\label{sec:dictionaries}

Programming with a read-only memory model can be tricky.
Even though, strictly speaking, an instruction
like \lstinline{x = y + 5} is interpreted as an assertion that the value
in the memory location assigned to \lstinline{x} is 5 more than the value in
the memory location assigned to \lstinline{y}, it also serves as a hint to
the Cairo runner as to the value that it ought to store in that location if the
program is to run to completion. Assignments to elements of an array can be
interpreted in the same way as long as conflicting values are never assigned
to the same element. In particular, Cairo code can act as though it is freely
appending values to the end of an array,
since the runner can easily arrange to put the corresponding values in memory.

In a Cairo program, however, it is often useful to make use of a dictionary of
key-value pairs in which values are read and reassigned. The CairoZero library
therefore includes a procedure, \lstinline{squash_dict}, that supports
such functionality. A dictionary is represented as an array of tuples of the
form (key, previous value, next value).
To assign a value to a key, the user appends a tuple with that key and next
value, relying on the runner to record the correct previous value.
To ``assign'' the value associated with a key to $x$, the user similarly
appends a tuple with both previous and next value asserted to be equal to $x$.
To ensure soundness, the user's program need only constrain the runner to
assign values that are consistent with a read-write history, which is to say,
for every tuple $(k, p, n)$ in the array, the next tuple $(k, p', n')$ satisfies
$n = p'$. In other words, when a key is set to some value, the next access
has to register that value as the previous value.

This is exactly what the procedure \lstinline{squash_dict} does:
given
pointers to the beginning and end of an array of triples $(k, p, n)$, it verifies that
the array has the property just described. It also returns a ``squashed''
version of the list of tuples, with a single entry for each key, giving the
initial and final values for that key. For example, suppose the initial array
is as follows:

\bigskip

\begin{center}
\begin{tabular}{| c | c | c |}
\hline
key & previous & next \\
\hline
\hline
7 & 3 & 2 \\
\hline
5 & 4 & 4 \\
\hline
7 & 2 & 10 \\
\hline
0 & 2 & 3 \\
\hline
7 & 10 & 0 \\
\hline
0 & 3 & 4 \\
\hline
0 & 4 & 5 \\
\hline
\end{tabular}
\end{center}

\bigskip

\noindent The \lstinline{squash_dict} function verifies that consecutive accesses
for each key are consistent and returns the following squashed version
of the dictionary:

\bigskip

\begin{center}
  \begin{tabular}{| c | c | c |}
  \hline
  key & previous & next \\
  \hline
  0 & 2 & 5 \\
  \hline
  5 & 4 & 4 \\
  \hline
  7 & 3 & 0 \\
  \hline
  \end{tabular}
\end{center}

\bigskip

\noindent More precisely, \lstinline{squash_dict} takes as input a pointer,
\lstinline{dict_acccesses}, to the beginning of the array of dictionary accesses,
a pointer, \lstinline{dict_accesses_end}, to the end of the array, and a pointer,
\lstinline{squashed_dict}, to the beginning of the squashed output dictionary.
It returns a pointer to the end of the squashed dictionary, and verifies,
first, that the input array is a consistent list of dictionary
accesses and, second, that the range of values between the input and output pointers
accurately describes the squashed version of the dictionary. The function
has the following signature:
\begin{lstlisting}
func squash_dict{range_check_ptr}(
  dict_accesses: DictAccess*, dict_accesses_end: DictAccess*,
  squashed_dict: DictAccess*
) -> (squashed_dict: DictAccess*)
\end{lstlisting}
It is implemented in about 200 lines of CairoZero code.
We have proved that the compiled code meets the following specification:
\begin{lstlisting}
def spec_squash_dict (mem : F → F) (κ : ℕ) (range_check_ptr : F)
    (dict_accesses dict_accesses_end squashed_dict : π_DictAccess mem)
    (ρ_range_check_ptr : F) (ρ_squashed_dict : π_DictAccess mem) : Prop :=
  κ < 2 ^ 50 →
  ∃ (n < κ) (m < κ),
    (dict_accesses_end : F) = dict_accesses + ↑(n * π_DictAccess.SIZE) ∧
    (ρ_squashed_dict : F) = squashed_dict + ↑(m * π_DictAccess.SIZE) ∧
    let squashed := π_array_to_list mem squashed_dict m,
        accesses := π_array_to_list mem dict_accesses n in
    -- The squashed list is sorted by key
    sorted_keys squashed ∧
    -- The set of keys is the same in the accesses and the squashed list
    (key_list squashed).to_finset = (key_list accesses).to_finset ∧
    -- every squashed entry is the squash of the key accesses for the
    -- squashed key.
    ∀ s : π_DictAccess mem, s ∈ squashed →
      is_squashed
        (accesses.filter (λ e : π_DictAccess mem, e.key = s.key)) s
\end{lstlisting}
In this specification, the Lean structure \lstinline{π_DictAccess} stores
a dictionary access $(k, p, n)$ together with its location
in memory, and the Lean constant \lstinline{π_DictAccess.SIZE} is equal to
3, the number of field elements needed to store $(k, p, n)$. Recall that
we use unicode characters like \lstinline{π} to ensure that autogenerated
names like these do not conflict with names of identifiers in the original
CairoZero code.

The function \lstinline{squash_dict} works by nondeterministically
guessing a sorted list of all the keys that are accessed in the input array,
and, for each key, a sorted list of entries in the initial array
that access the key. In more prosaic
terms, this means that the CairoZero code provides hints to the runner to put
these lists in memory and then makes sufficient assertions about those
memory locations to verify that the copy meets that specification.
One can verify the claims about the list of keys and each list of accesses
associated to each key
by looping through them and checking.
The existence of this data implies that there is an injective function from the
union of the entries
corresponding to all the keys to the entries in the input array, and
asserting that the total number of entries associated to the keys is equal
to the total number of entries in the input array implies that the function
is a bijection.

Recall from Section~\ref{sec:memory:management:and:range:checks} that the
parameter \lstinline{κ} in \lstinline{spec_squash_dict} represents the number
of steps in the
execution trace. The specification assumes that
\lstinline{κ < 2 ^ 50}, a bound so large that the condition
always holds in practice. That condition can moreover be verified in each
case from the STARK certificate.
Our proof of \lstinline{spec_squash_dict} provides a nice example of
why such a bound is sometimes needed.
Proving the correctness of the CairoZero calculation that establishes that the total
number of entries associated to the keys is equal
to the total number of entries in the input array
requires showing (among other things)
that no entry in the input array is counted twice. The CairoZero code ensures this
by setting
bounds on the memory locations of the entries, but, crucially, our proof must
also show that the lists of entry locations does not wrap around the finite
field to reuse the same entry.
The fact that we can show that the value of \lstinline{n}
asserted to exist
in the specification is bounded by \lstinline{κ} provides that guarantee.

One can also nondeterministically guess the squashed output
dictionary and check that it has the right size, namely, the number of keys.
While looping through the list of accesses associated to a key $k$, one can check
that the entry $(k, p, n)$ in the squashed dictionary is correct by ensuring
that $p$ agrees with the first triple in the list of accesses associated to $k$
and that $n$ agrees with the last triple in that list.
We have verified that, taken together,
all these checks imply that the specification holds. Stating all the
invariants correctly and handling corner cases was a substantial effort,
and the file \lstinline{squast_dict_spec.lean} is more than 1,500 lines long.

\section{Methodology}
\label{sec:methodology}

We have reported on the means we have developed to deal with quirks of the
Cairo architecture that stem from the need to encode Cairo computations
efficiently in a STARK certificate. Beyond that, many of the methods we have used are
routine for software verification. But some aspects
of the way we have implemented these methods are notable,
since they have enabled us to put the methods to use in a production setting.
In this section, we discuss
some of the pragmatic choices we have made and assess their effectiveness.

It is notable that our end-to-end correctness proofs are carried out
within a single foundational proof assistant. Systems such as Why3 \cite{filliatre:paskevich:13},
Dafny \cite{leino:10}, and $F^*$ \cite{swamy:et:al:13} extract verification conditions from imperative programs,
but they do not generally verify those conditions with respect to a mathematical specification of
a machine model. These systems also tend to rely on automation, like SMT solvers,
that has to be trusted.
In contrast, all our theorems are stated in the context of a precise axiomatic
foundation and the proofs are checked by a small trusted kernel, for which
independent reference checkers are available.
This provides a high degree of confidence that the machine code meets its high-level specifications.
Similar approaches to verifying code with respect to machine semantics include
MM0 \cite{carneiro:20} and work by Myreen, Slind, and Gordon \cite{myreen:et:al:09}.

Another advantage of embedding the verification in a foundational
proof assistant is that the availability of an ambient mathematical library \cite{mathlib}
means that we can make use of any mathematical concepts that are needed to make sense
of the high-level specification.
Our verification of the digital signature recovery algorithm required reasoning
about elliptic curves, as well as dealing with bounds and casts of integers to a
finite field. We were able to carry out this reasoning in the same proof assistant
that we used to carry out low-level reasoning about the machine code.

It is notable that the development of our tooling did not hamper the development of the
CairoZero compiler or its library.
When we began our project, the compiler was already being used in production,
and it was under continuous development. Requesting substantial
changes to the compiler code base would have slowed our efforts,
requiring not only coordination with the compiler team but also
extensive code review. With our approach, we were able to work under the radar,
harvesting just enough data from the compiler for us to construct our proofs.
For example, we found that justifying equality assertions between compound
terms did not require a detailed understanding of the process by which the
compiler carried out the calculations; it was enough to simply keep track of
the intermediate assertions and pass those equations to Lean's simplifier.

An alternative approach to end-to-end verification is to verify a compiler with
respect to a deeply embedded semantics. This is the approach taken by
CompCert \cite{leroy:09}, CakeML~\cite{kumar:et:al:14,abrahammson:et:al:20}, and the Bedrock project (e.g.~\cite{chlipala:13}).
But the CairoZero language was constantly evolving and there is still no formal specification of
its semantics, even though the meaning of a CairoZero program is
intuitively clear in general. Our approach gives us the freedom
to generate specifications with confidence that
they are correct, since they are backed up by formal proof.
Producing proofs of correctness at compile time
avoids having to model parts of the compiler that are irrelevant to correctness,\footnote{We are grateful to a reviewer for pointing out that at least one component of CompCert
uses this strategy;
see the discussion in \cite[Section 2.2]{leroy:09} and
the discussion of register allocation in \cite[Section4.2]{leroy:09}.}
and it does not require us to find a clean separation between those parts
and the ones that are. It also
avoids the need to verify behaviors that don't arise in practice. For example,
CairoZero allows for arbitrary labels and jumps, and programmers are free to write
whatever spaghetti code they want. Our verification tooling is designed to
work on regular control flow graphs, and will simply fail otherwise.
This leaves the decision with CairoZero developers as to whether to revise
their CairoZero code to fit our verification model, to verify their code by
hand, or to leave it formally unverified. Thus our approach provides tools that
are effective in practice without dictating or constraining the language development.

Perhaps most striking is our decision to construct
correctness proofs by generating Lean source code that is then elaborated
and checked by the same Lean process that elaborates and checks hand-written proofs. This means that our tool automatically constructs long, complex proofs in a system that has been carefully designed to support synergetic user
interaction. This may seem odd and counterproductive. But we found that the
compilation process is deterministic enough to make it possible to construct
these proofs, and that we could make use of similarly deterministic and predictable automation in Lean.

Moreover, we found that the approach supports an efficient workflow for the
developers of the verification tool as well as for users of the tool who wish
to verify their CairoZero specifications.
To verify a CairoZero program, a user runs our verification tool on the main file,
which can import other CairoZero program files. Our tool calls the compiler to
compile the program and then generates a Lean description of the compiled
code, a Lean specification file for each CairoZero source file, and proofs that
the compiled code meets the specifications. By convention, the specification
files, which are typically the only formal content the user needs to inspect
and modify, are kept with the source files. For example, a CairoZero file {\tt foo.cairo} gives rise
to a specification file \verb~foo_spec.lean~ in the same directory.
The remaining files are kept in a {\tt verification} folder in the directory
containing the main CairoZero source file. Compiling the files immediately after
the tool is run confirms that the compiled code meets the autogenerated
default specifications. Compiling them again after the user adds their own specifications
to the spec files and proves that they follow from the autogenerated specifications
ensures that the code meets their specifications. The correctness proofs do
nothing more than apply the theorem that says that the autogenerated
specification implies the user one, so if the initial correctness proofs have
already been checked and Lean accepts the proofs in the user specification file,
it is unlikely that the subsequent check of the correctness proofs will fail.
For the application described in Section~\ref{sec:implementing:secp},
compiling and checking all the files in Lean requires only a few minutes
on an ordinary desktop. Compiling and checking the specification files alone,
with the user specifications and correctness proofs, takes less than two minutes from scratch. In practice, the user files are checked incrementally in real time, as the user types them into the editor.

This results in a congenial workflow. After first running the verification tool,
a user can compile the files in the verification folder to confirm that the
specifications are well formed and the correctness proofs are valid. The user can then focus on writing the
specifications and proving them correct. We have arranged it so that if
the verification tool is run again in the presence of existing specification files,
we do not overwrite any of the user-supplied content. We only add or change the autogenerated
specifications, as well as the arguments to the specifications when the
arguments to the corresponding CairoZero functions change. (The tool leaves
comments in the file so that the user can see what has changed.) That way,
when the CairoZero code changes, the user only needs to make corresponding changes
to the specifications. Moreover, when verifying another CairoZero file with
overlapping dependencies, one can make use of the same specifications.
This has made it possible for us to verify the CairoZero library
one step at a time.

Our approach has also had important benefits for the development of the
verification tool. We started our project by compiling simple
programs, extracting Lean descriptions of the compiled code, and
writing and proving specifications by hand. This helped us determine
what the autogenerated specifications should look like and taught us how to construct
the correctness proofs. We then simply had to write Python code that
did the same thing automatically.
We were able to iteratively extend the tool
to handle other aspects of the CairoZero language: if-then-else blocks,
recursive calls, structures, loops, and so on. As we worked through files
in the CairoZero library, whenever we came across a feature the verifier was
not equipped to handle, we could figure out how to handle the feature manually,
and then extend the tool to handle that and future instances. Debugging was
similarly straightforward: whenever one of our autogenerated proofs failed,
we could open the file, go to the error, and use Lean's rich editor interface
to inspect the proof state.
Once we figured out how to repair the error manually, it was generally not hard to modify
the verification tool to produce the desired behavior automatically.
Our generated code is structured, commented, and readable. It slightly more
verbose and formulaic than proofs one would write by hand, but it is otherwise similar.

In sum, formal verification requires
a synergetic combination of automation and user interaction.
One of our most important findings is that using automation to generate
formal content that can be inspected and modified interactively is a remarkably
powerful and effective means to that end.

\section{Conclusions and Future Work}
\label{sec:conclusions}

We have presented a means of verifying soundness of programs
written in the CairoZero language with respect to a low-level machine model,
and we have demonstrated its practical use by verifying
CairoZero library procedures for computation with the secp256k1 and secp256r1
elliptic curves and for validating cryptographic signatures.
We have similarly verified other fundamental components of the CairoZero library,
including a procedure that CairoZero programmers can use to simulate the behavior of
read-write dictionaries in Cairo's read-only memory model.
Cairo programs are used extensively to carry out financial transactions,
and they are carefully optimized to reduce the cost of
on-chain verification, resulting in more complicated code.
Having workable means of verifying their correctness
is therefore essential.

In Section~\ref{sec:methodology}, we have already cited some other approaches toward verifying
a functional specification down to machine code, and in Section~\ref{sec:implementing:secp},
we cited various formalizations of the associativity of the group law for elliptic curves.
In recent years, there has been extensive work on verification of
cryptographic primitives \cite{erbsen:20,portzenko:et:al:20,schwabe:et:al:21,zinzindohoue:et:al:17}, including the kind of digital signature recovery
described here.
As we have explained, however, verification of CairoZero programs requires
dealing with specific features of the language and machine model,
and it is notable that we have achieved end-to-end verification in a
foundational proof assistant.

Our current verification efforts are now focused on the next generation
programming language, now called ``Cairo,'' which is implemented in Rust.
The move from CairoZero to Cairo roughly corresponded to the Lean community's move from Lean 3 to Lean 4.
We have not found a need to update our CairoZero verification tool to output
Lean 4 code, but we have ported the entire Cairo semantics, as well as our
stepping tactics, to Lean 4. We have also modified a similar verification tool
for Cairo to work with Lean 4, and the changes were straightforward.

The new Cairo provides typing
mechanisms that provide guarantees that programs written in the language terminate
successfully (although possibly by raising an exception), which is to say, the runner will be able to produce a certificate that shows that the program reaches
a final state with a specified semantics. For that purpose, it is important to
verify the \emph{completeness} of basic library functions as well as their
soundness, which raises new challenges.

The methods we have developed here, however,
are still central to our approach.
In particular, we still automatically generate and verify the correctness of naive high-level specifications,
leaving the user with the more interesting task of reducing the desired specifications
to those.
We will report on these efforts in a future paper.

\paragraph{Acknowledgements}
We are grateful to two anonymous reviewers for many helpful comments, suggestions, and corrections.

\bibliographystyle{plain}
\bibliography{verifying_compiler_arxiv}

\end{document}